\documentclass[a4paper]{jpconf}
\usepackage{graphicx}
\begin{document}
\title{Atmospheric Neutrinos}

\author{Thomas K. Gaisser}

\address{Bartol Research Institute and Department of Physics and Astronomy\\
University of Delaware, Newark, DE, 19716, USA}

\ead{gaisser@bartol.udel.edu}

\begin{abstract}
In view of the observation by IceCube of high-energy astrophysical neutrinos, it
is important to quantify the uncertainty in the background of atmospheric neutrinos.
There are two sources of uncertainty, the imperfect knowledge of the spectrum and
composition of the primary cosmic rays that produce the neutrinos and 
the limited understanding of hadron production, including charm, at high energy.
This paper is an overview of both aspects.

\end{abstract}

\section{Introduction}

To calculate the spectrum of atmospheric neutrinos up to the PeV range it is
necessary to use as input the primary cosmic-ray spectrum to $\sim 100$ PeV,
well into the energy region where data is available only from ground-based air shower
experiments.  These experiments measure the spectrum of energy per particle.  The
relative fraction of various nuclei is obtained indirectly from ground-level observables,
such as the ratio of electrons to electromagnetic particles, or by comparing the
depth of shower maximum to expectation.  In both cases the results depend on simulations
with event generators that embody the hadronic physics. 
The composition is important because it is the energy per nucleon that is relevant
for the atmospheric neutrino spectrum, rather than the all-particle spectrum measured
by the air shower experiments.  

Models of the spectrum and composition are anchored to direct measurements which 
extend to about 100 TeV for protons and helium, but an order of magnitude lower
for the carbon group~\cite{Seo:2012pw}.  Fits to direct measurements need
to be extrapolated to higher energy and joined to results from air shower
experiments, such as KASCADE~\cite{Antoni:2005wq}.  The classic example of
such a parameterization is the Polygonato Model of H\"{o}randel~\cite{Hoerandel:2002yg}.
Understanding the composition and spectrum through the knee region
where the spectrum steepens significantly is crucial for calculating
the atmospheric neutrino spectrum above 100~TeV.

\begin{figure}[!t]
\includegraphics[width=3.4in]{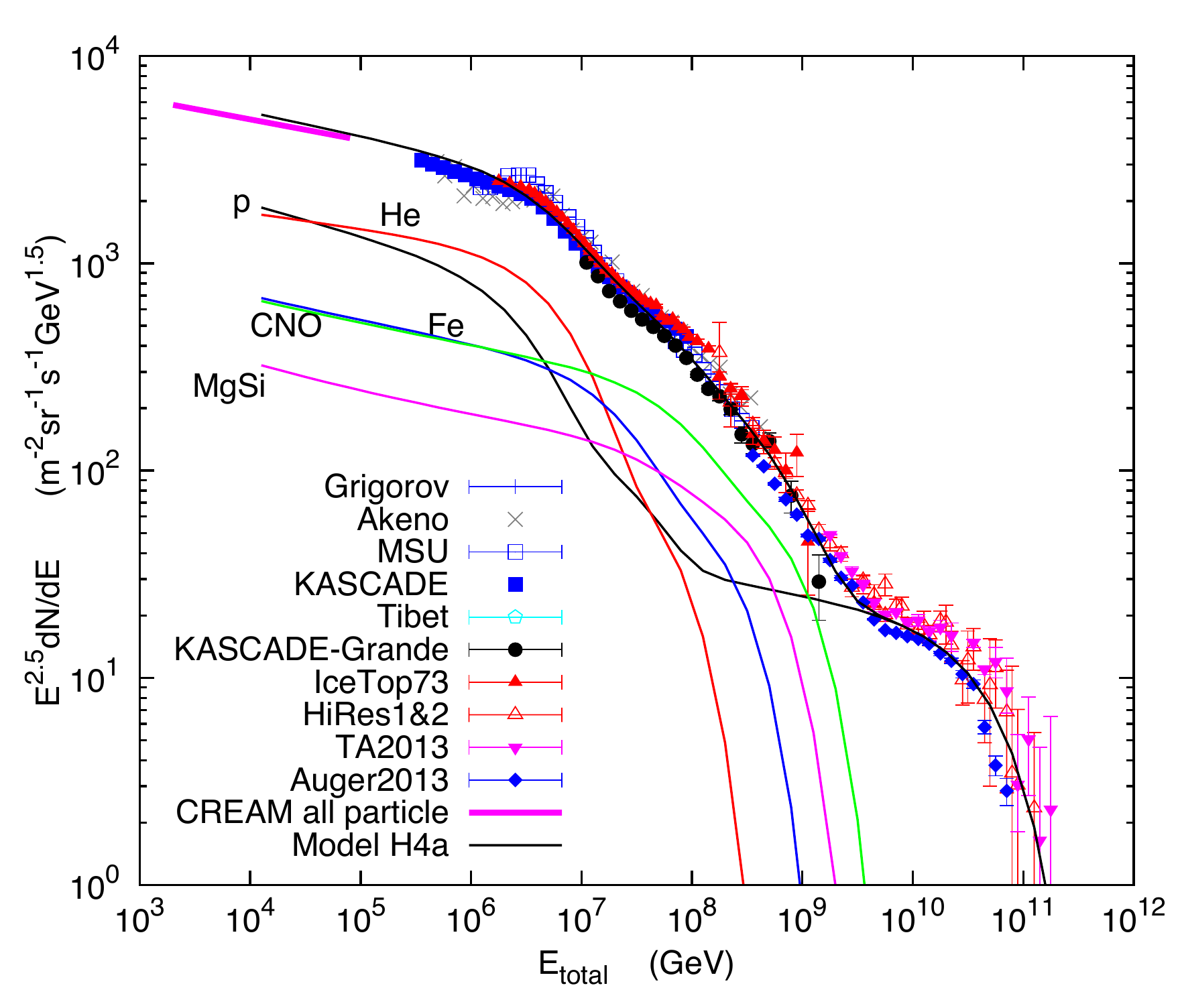}\includegraphics[width=2.9in]{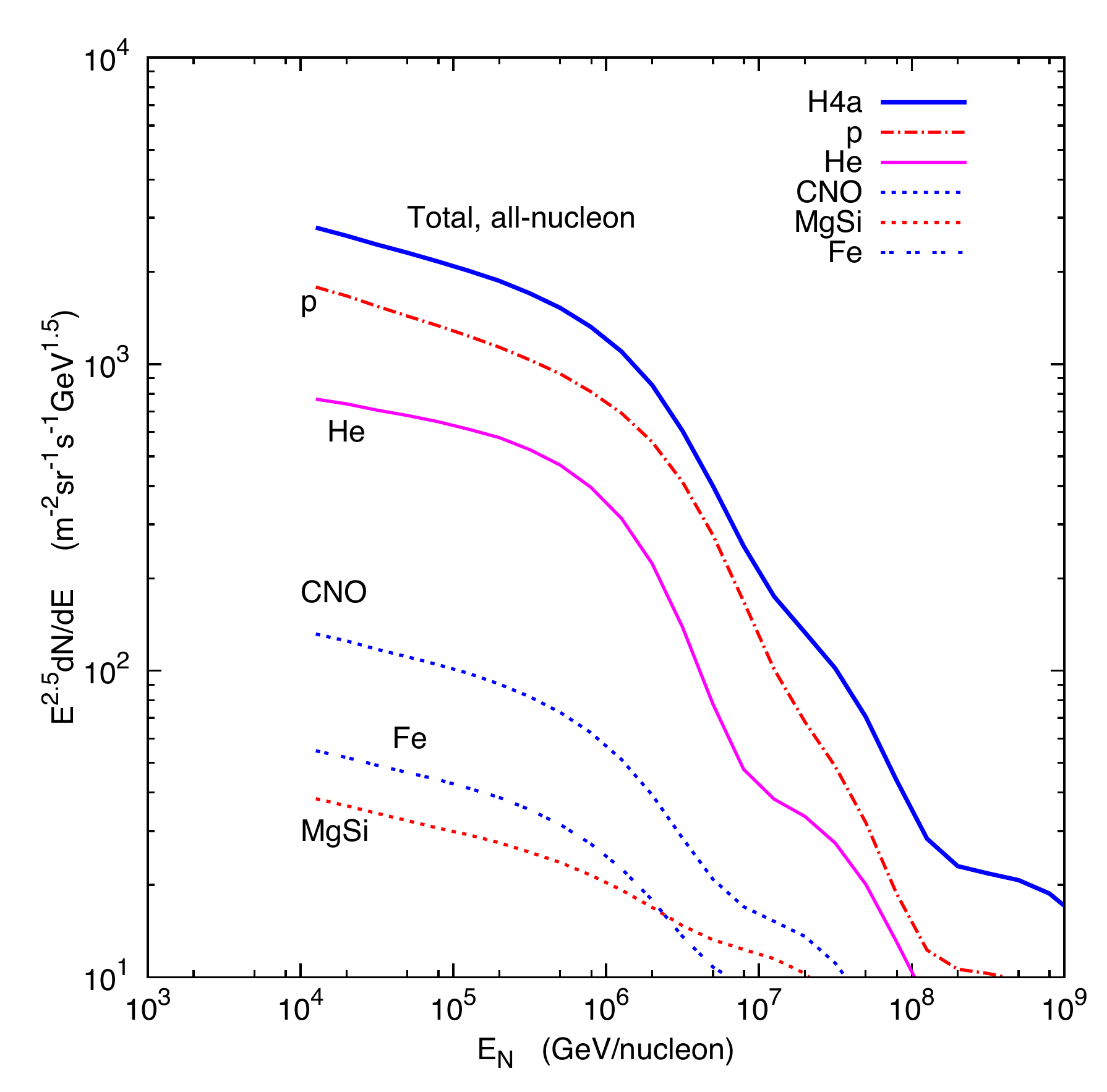}
\caption{Cosmic ray energy spectrum of nuclei (left) and the corresponding spectrum of nucleons (right)}
\label{taup2013}
\end{figure}

These points are illustrated in the first two figures of a paper~\cite{Gaisser:2015ana}
from TAUP 2013, reproduced here in Fig.~\ref{taup2013}.  Around $30$~TeV, the helium contribution, which has a
slightly harder spectrum, crosses the proton contribution to the all-particle
spectrum.  At the same energy per nucleon, however, 
the contribution of the four nucleons
in He to the spectrum of nucleons is about a factor of three below 
protons.  


\begin{figure}[!htb]
\includegraphics[width=0.6\textwidth]{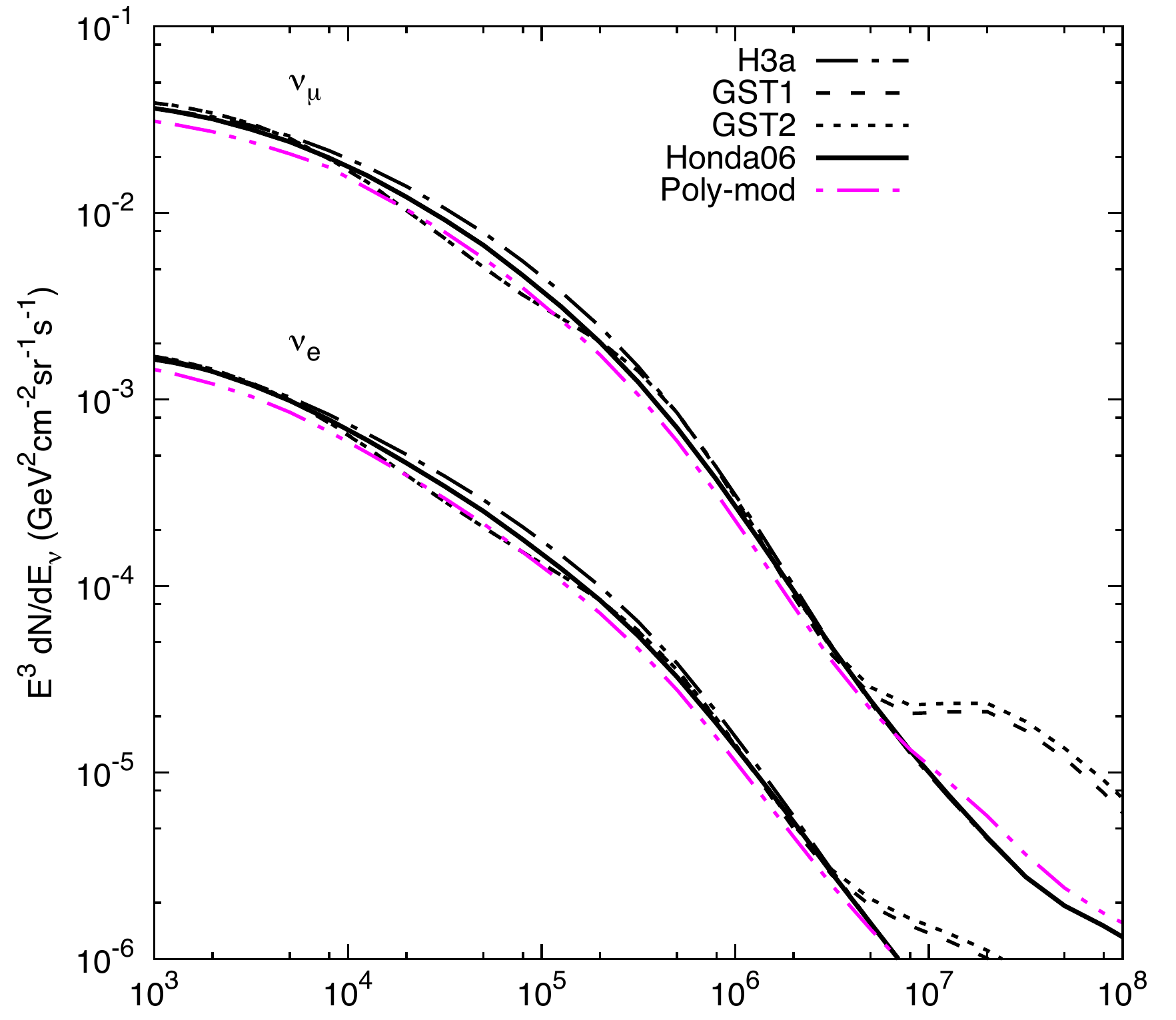}\\
\includegraphics[width=0.6\textwidth]{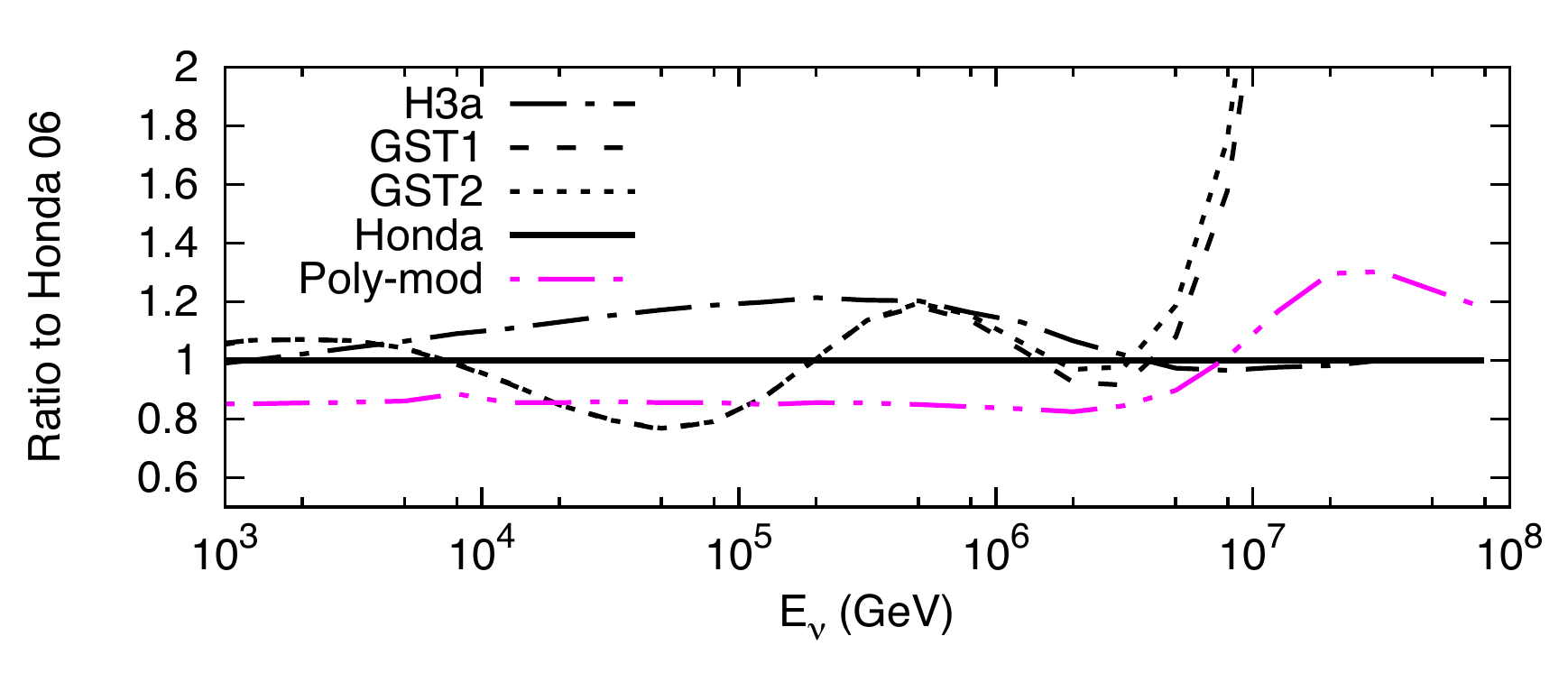}
\caption{Top panel: Conventional $\nu+\bar{\nu}$ fluxes 
averaged over all zenith angles for 5 primary spectra calculated with 
the EPOS-LHC model~\cite{Pierog:2013ria}; Bottom panel: ratio to
Honda~06.
\label{fig:primary}
}
\end{figure}

The plan of this paper is to compare the atmospheric neutrino spectrum
using a reference hadronic interaction model with several different models
of the primary spectrum in the next section.  The following section takes
a single model of the primary spectrum and shows the atmospheric neutrino spectrum
for several hadronic event generators.  Prompt neutrinos and charm are discussed
separately in the last section.

\section{Models of the primary spectrum}

The uncertainty in the flux of conventional neutrinos from decay of $K^\pm$ and $\pi^\pm$
is illustrated in Fig.~\ref{fig:primary} by comparing the results of five different models 
of the primary spectrum and composition.  All are calculated numerically as 
described in Ref.~\cite{Gaisser:2014eaa} using the same hadronic interaction model. 

The spectrum models 
described and tabulated in Ref.~\cite{Gaisser:2013bla} are designed
to connect with direct measurements below $100$~TeV, to describe the knee region, 
and also to be applicable at the highest energies.  The common feature of these
models is that they consist of several populations of particles, with successively
higher cutoffs in rigidity.    In the H3a model~\cite{Gaisser:2012zz}, for example,
each population consists of the five major nuclear
groups, represented by p, He, N, Si and Fe.
The first population has an exponential cutoff at $R_c=4$~PV with the spectral index
of each mass group tuned to match direct measurements~(e.g.~\cite{Seo:2012pw}).  In particular,
helium has a harder spectrum in the first population than protons.  The second population
with $R_c=30$~PV is a Galactic component of unknown origin that fills the gap that
would otherwise occur below the assumed extra-galactic population~\cite{Hillas:2005cs}.
In H3a the extra-galactic population contains all nuclei with a cutoff of $R_c=2$~EV.
In the two models introduced in Ref.~\cite{Gaisser:2013bla} (GST1 and GST2), the populations
above the knee contain no intermediate mass nuclei.  As a consequence, the spectrum of
nucleons is dominated by protons above $\sim 10^7$~GeV.  This feature in the composition
of the GST models is reflected in the corresponding neutrino spectra  
above a few PeV (Fig.~\ref{fig:primary}).

The primary spectrum used for two benchmark calculations of the atmospheric neutrino flux 
Bartol~\cite{Barr:2004br} and Honda~\cite{Honda:2006qj} is the high-helium version of a spectrum fit given
in Ref.~\cite{Gaisser:2001jw}.  That parameterization was made for energy per nucleon less than $\sim 100$~TeV.
Here, populations 2 and 3 of H3a are used to extend this model (Honda06) through the knee region and beyond.
The detailed Polygonato model~\cite{Hoerandel:2002yg,Hoerandel:2004gv} provides an extrapolation
specifically designed to describe the knee in the cosmic-ray spectrum in the PeV energy range,
but not for the much higher EeV energy range.  Population 3 of H3a is used here to extend the
version of Polygonato with a rigidity-dependent cutoff in the knee region to higher energy 
(``Poly-mod" in Fig.~\ref{fig:primary}).
In both cases, the H3a components are renormalized to make a smooth transition.

\begin{figure}[!htb]
\includegraphics[width=0.6\textwidth]{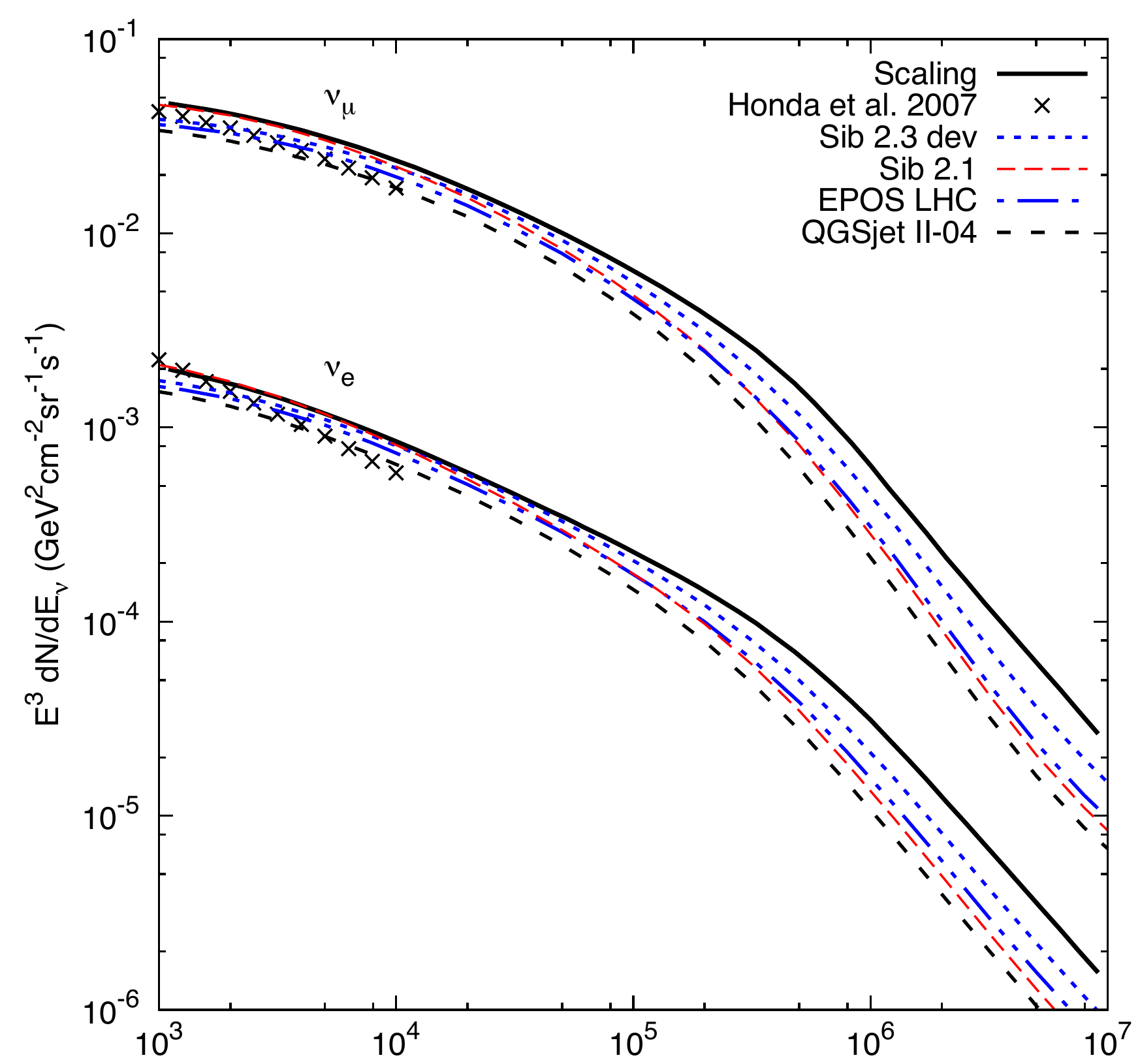}\\
\includegraphics[width=0.6\textwidth]{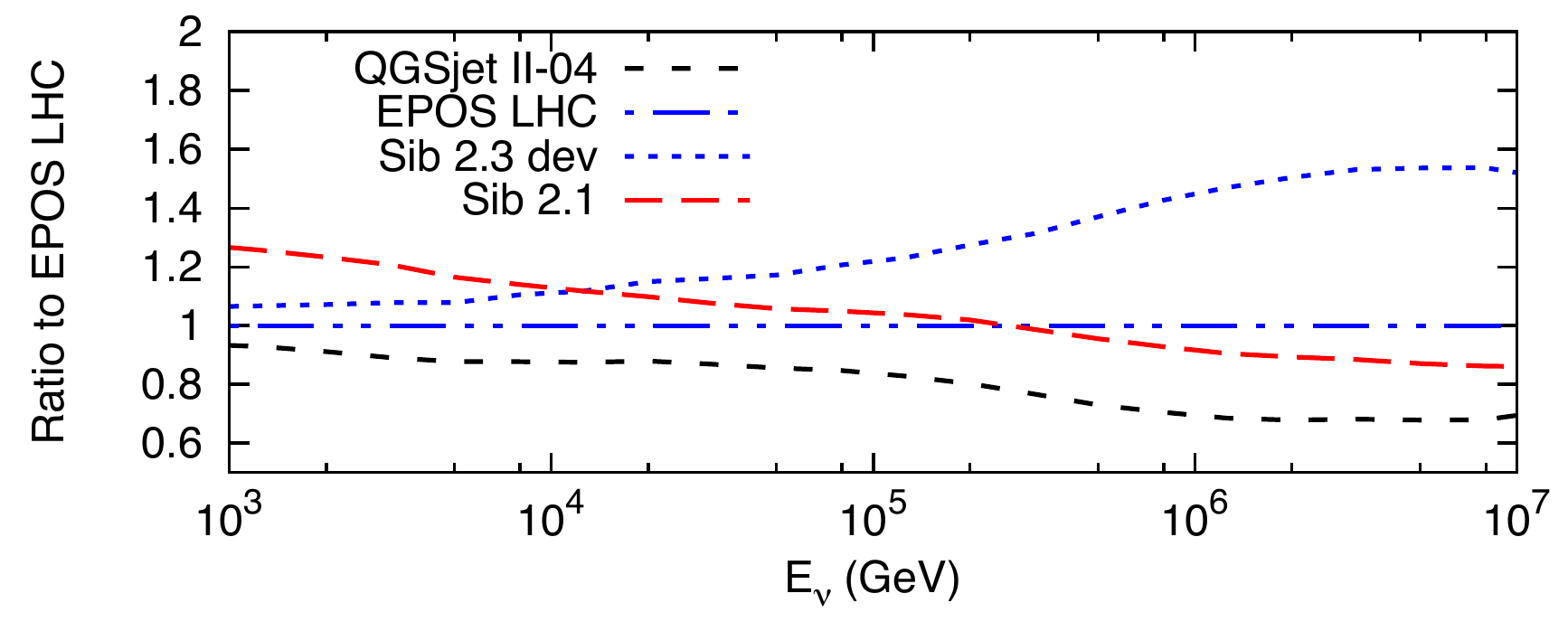}
\caption{Top panel: Conventional neutrino plus anti-neutrino fluxes 
averaged over all zenith angles for several interaction models;
Bottom panel: Ratios to fluxes calculated with EPOS-LHC.  (Note: The
numerical calculations shown here by the lines do not include neutrinos
from decay of muons, which are negligible for $\nu_\mu$ with $E>$~TeV.
There is still some contribution to TeV $\nu_e$, which probably accounts
for the steeper $\nu_e$ flux from the tables of Ref.~\cite{Honda:2006qj}, which
do include this contribution.)
\label{fig:hadronic}
}
\end{figure}

\section{Hadronic interaction models} 

Spectrum-weighted moments have been tabulated as a function of beam energy 
for several interaction models
using the CRMC program~\cite{Pierog:2015xyz}.  The energy-dependent inclusive cross
sections are then derived for each model and the neutrino fluxes are calculated
using the numerical method of Ref.~\cite{Gondolo:1995fq} as described in Ref.~\cite{Gaisser:2014eaa}.  
The hadronic interaction models are listed in Table~\ref{tab:hadronic}.
\begin{table}[h]
\begin{center}
\begin{tabular}{l|l}  \hline
Model & Comment \\ \hline  \hline
Scaling & Energy-independent Z-factors from ~\cite{Gaisser:1990vg} \\ \hline
Honda~\cite{Sanuki:2006yd} & Tuned to atmospheric $\mu^\pm$; used in~\cite{Honda:2006qj} for atmospheric $\nu$ \\ \hline
QGSjet II-04~\cite{Ostapchenko:2013pia} & Post-LHC version of QGSjet  \\ \hline
EPOS LHC~\cite{Pierog:2013ria} & Post-LHC version of EPOS  \\ \hline
Sib 2.3 dev~\cite{Riehn:2015oba} & Development version of post-LHC SIBYLL  \\ \hline
Sib 2.1~\cite{Ahn:2009wx} & Pre-LHC Sibyll \\ \hline
\end{tabular}
\caption{List of hadronic interaction models for the atmospheric neutrino 
spectra in Fig.~\ref{fig:hadronic}.
\label{tab:hadronic}}

\end{center}
\end{table}
The neutrino fluxes from these models are compared to each other in Fig.~\ref{fig:hadronic}.

\begin{figure}[!htb]
\includegraphics[width=0.6\textwidth]{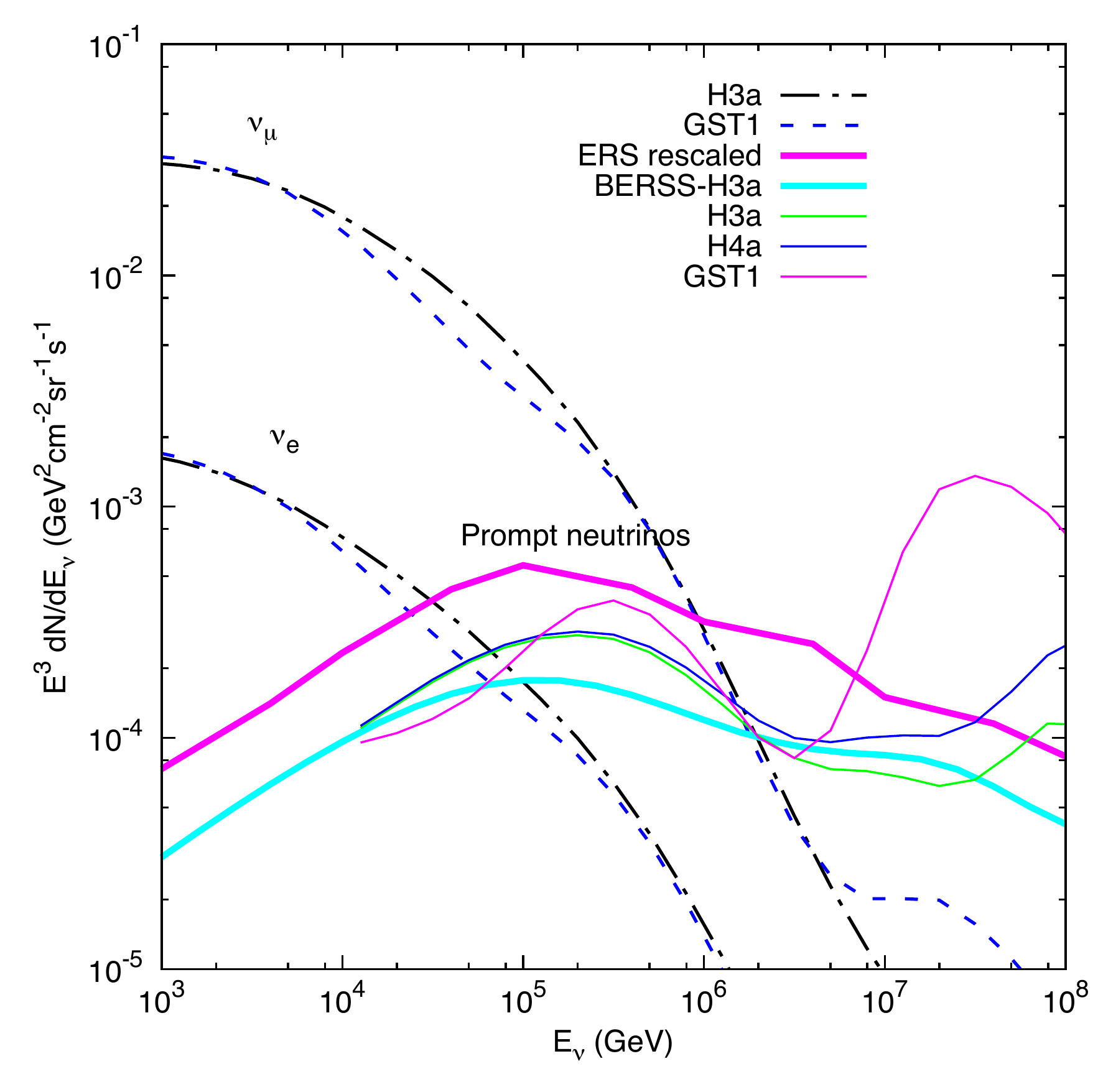}\\
\caption{Various estimates of the intensity of prompt atmospheric neutrinos
from decay of charm compared to conventional atmospheric neutrinos.
\label{fig:prompt}
}
\end{figure}

\section{Prompt neutrinos from decay of charm}

There are many calculations of the flux of prompt neutrinos from
decay of charmed hadrons produced by interactions of cosmic rays in the
atmosphere.  Several of these are shown in Fig.~\ref{fig:prompt} with 
solid lines.  There are uncertainties in the level of prompt neutrinos
related to limited knowledge of production of charmed hadrons and also
from the primary spectrum and composition.  Examples of both are
shown in the figure.  The heavy solid lines are from ERS~\cite{Enberg:2008te} (rescaled
to include the effect of the knee) and BERSS~\cite{Bhattacharya:2015jpa}, both
assuming the H3a model for the primary spectrum and composition.

To illustrate the effect of differences in primary spectrum,
a simple model of the energy-dependent charm production was
made based on Fig.~2 of BERSS~\cite{Bhattacharya:2015jpa}.  The prompt flux was then
calculated using the same numerical method as for the conventional
neutrinos, taking account of the much higher critical energies
for charmed hadrons ($3.84\times 10^7$~GeV for $D^\pm$ and 
$9.71\times 10^7$~GeV for $D^0$).  The results are shown for
spectrum models H3a, H4a and GST1 by the three thin continuous
lines in the figure.  (H4a differs from H3a only in the 3rd
(extragalactic) population, which is all protons for H4a.)
For comparison, the fluxes of conventional atmospheric neutrinos
are shown by the broken lines for H3a and GST1.

Fluxes of $\nu_e$ and $\nu_\mu$ from decay of charm are nearly equal, unlike
the case for conventional neutrinos, for which $\nu_e / \nu_\mu\approx 0.04$.
For the examples shown here, the crossover energy at which the prompt neutrino
flux becomes larger than the conventional is between 10 and 100~TeV
for $\nu_e$ and between 1 and 3~PeV for muon neutrinos.  The effect of the
protons in the GST models is again apparent in the increased intensity above $10$~PeV.
Sibyll 2.3~\cite{Riehn:2015oba} includes production of charm and gives a
prompt flux within the range shown in Fig.~\ref{fig:prompt}.

A new calculation of prompt neutrinos by Garzelli et al.~\cite{Garzelli:2015psa} 
that starts from next to leading
order QCD for hadro-production of charm gives results that are similar in magnitude
to Fig.~\ref{fig:prompt}.  Reference~\cite{Garzelli:2015psa} (also presented at TAUP~2015)
includes an extensive evaluation of uncertainties in the flux of neutrinos from decay of charm. 
Taken together, the recent calculations are in broad
agreement and predict levels of prompt neutrinos within a factor of two
of the results shown here.

\section[*]{Acknowledgments}
I thank Felipe Campos Penha for helpful discussions of lepton fluxes.
I thank Hans Dembinski for providing the Z-factors for
the various hadronic interaction event generators used in this paper,
and I thank Ralph Engel, Anatoli Fedynitch, Felix Riehn 
and Todor Stanev for helpful
discussion on SIBYLL and related topics.
This research is supported in part by
grants from the U.S. National Science Foundation, NSF-PHY-1205809
and from the U.S. Department of Energy, 12ER41808.
\vspace{0.2cm}\eject

\section[*]{References}
\vspace{0.2cm}

\bibliographystyle{iopart-num}
\bibliography{Gaisser}

\end{document}